\documentclass{amsart}

\usepackage[table,xcdraw]{xcolor}
\theoremstyle{definition}

\theoremstyle{remark}

\usepackage{kotex}

\numberwithin{equation}{section}
\usepackage{url}
\usepackage{amssymb}
\usepackage{graphicx} 
\usepackage{subcaption}
\usepackage{hyperref}
\usepackage{listings}
\usepackage{xcolor} 
\graphicspath{ {/Users/ekim/Desktop/Stuff} }
\usepackage{geometry}
\begin{document}
	
	\title{Factors Involved in Cancer Screening Participation: Multilevel Mediation Model}
	
	\author{Donghyun (Ethan) Kim}
	\address{Gyeonggi Suwon International School, 451 YeongTong-Ro, YeongTong-Gu, Suwon-Si, Gyeonggi-Do, Republic of Korea}
	
	\email{ethank11k@gmail.com}

	

	\dedicatory{{\normalfont Gyeonggi Suwon International School, 451 YeongTong-Ro, YeongTong-Gu, Suwon-Si, Gyeonggi-Do, Republic of Korea}}

	\begin{abstract}
		In this paper, we identify the factors associated with cancer screening participation in Korea. We expand upon previous studies through a multilevel mediation model and a composite regional socioeconomic status index which combines education level and income level. Results of the model indicate that education level, nutritional education status and income level are significantly associated with cancer screening participation. With our findings in mind, we recommend health authorities to increase promotional health campaigns toward certain at-risk groups and expand the availability of nutrition education programs.   
	\end{abstract}
	
	\maketitle
	
	\section{Introduction} 
	
	Cancer is the leading cause of death in Korea. In 2018, for every 100,000 people, cancer was responsible for 154.3 deaths (26.5\% of all deaths), a 0.2\% increase from 2017 (\cite{K}).
	
	Early detection and diagnosis of cancers significantly increases a patient's chance of survival and reduces the costs associated with treatment. A study in the United States found that women diagnosed with cervical cancer had significantly greater 5 year survival rates when the cancer was detected in early stages compared to diagnosis at advanced stages (92\% vs 17\%) (\cite{M}).
	
	Korea's National Cancer Screening Program covers 6 major cancers, namely gastric cancer, colorectal cancer, breast cancer, cervical cancer, liver cancer, and lung cancer. For National Health Insurance (NHI) beneficiaries in the lower 50\% income bracket, cancer screening tests are completely free of cost. For NHI beneficiaries in the upper 50\% income bracket, 90\% of cancer screening costs are covered by the NHI and 10\% is left for the individual to pay (\cite{N}). 
	
	Even with the government's efforts to increase the availability of cancer screening tests, the participation rate was 55.6 \% in 2019; rates did not increase much from 50.1 \% in 2015 (\cite{L}). In this study, we aim to expand upon previous research and identify the factors associated with cancer screening participation with a multilevel mediation model.
	
	First, section 2 will discuss factors found from previous studies. Next, section 3 will present the model with some preliminary statistical tests. Finally, sections 4 and 5 will discuss the results and recommend a number of actions moving forward.
	
	\section{Factors Involved}
	
	\subsection{General Factors from Korean Study and other papers}
	
	We will first take a look at some previous studies on the factors involved in cancer screening tests. These factors will serve as a foundation for the model presented in this paper.
	
	In 2012, authors Shin, Ji-Yeon and Lee, Duk-Hee conducted a study on the factors associated with the use of gastric cancer screening services in Korea. Through multivariate analysis, they found that education level, alcohol consumption, and smoking habits are significantly associated with participation in gastric cancer screening. More specifically, individuals who graduated from middle school and high school (adjusted Odds Ratio = 1.21) and those who graduated from college or higher (adjusted Odds Ratio = 1.40) were much more likely to utilize gastric cancer screening services than those of lower education level. Binge drinkers (adjusted Odds Ratio = 0.94) and frequent binge drinkers (adjusted Odds Ratio = 0.74) were less likely to utilize gastric cancer screening services than non-binge drinkers. Current smokers (adjusted Odds Ratio = 0.75) and ex-smokers (adjusted Odds Ratio = 0.93) were less likely to utilize gastric cancer screening services as well.
	
	Monthly household income, however, was found to be positively associated with the use of gastric cancer screening services (4th quartile: adjusted Odds Ratio = 1.31; 3rd quartile: adjusted Odds Ratio = 1.09).
	
	The study also found that dietary habits ("trying to eat more vegetables") are positively associated with cancer screening participation when uni-variate logistic regression was conducted (Very often: Odds Ratio = 1.00; Somewhat often: Odds Ratio = 0.98; Not at all: Odds Ratio = 0.80) (\cite{A}).  
	
	Another study from Korea in 2016 found that moderate level of physical activity (1-4 times a week) is positively associated with cancer screening participation (Odds Ratio = 1.20 for organized screening and Odds Ratio = 1.41 for opportunistic screening) (\cite{B}). 
	
	\subsection{Relations between factors}
	
	In this section, we will take a look at previous studies that examined the relationship between the aforementioned factors.
	
	Authors Catherine E. Ross and Chia-ling Wu concluded in their paper that individuals with higher education levels are more likely to exercise (unstandardized regression coefficient: .295), less likely to smoke (unstandardized regression coefficient: -.046), and more likely to drink heavily (unstandardized regression coefficient: .036) (\cite{C}). In another study, authors David M. Cutler and Adriana Lleras-Muney found that education is positively associated with one's diet (regression coefficient: 0.0658) (\cite{D}). $\\$
	
	Looking at income, author M. Christopher Auld found that drinking is associated with higher income (10\% higher for moderate drinking and 12\% higher for heavy drinking) and smoking is associated with lower income (smokers earn 8\% less than those who do not smoke) (\cite{E}). Another study found that the quality of an individual's diet/nutrition improved with higher income levels (\cite{F}). Finally, authors Brad Humphreys and Jane Ruseski found positive associations between exercise (including swimming, golfing, weight lifting, and running) and income levels (\cite{G}). 
	
	\subsection{Socioeconomic Status Index}
	
	Regional and individual socioeconomic status indices are often based on income, assets, occupation, and education (or a combination of these variables). These indices have varied uses depending on the study in question, including measuring inequality or acting as a multilevel predictor in a model. An example of a composite socioeconomic status index is the "Kuppuswamy socioeconomic scale," which combines assets, education level, occupation, and income (\cite{H}).

	\section{Method}
	
	\subsection{Participants}
	
	This study is based on data obtained from the Seventh  Korea National Health and Nutrition Examination Survey, which includes data from the years 2016 to 2018. This particular dataset includes information from 10611 households (24269 individuals) and consists of a survey of health and nutrition and a health examination. Study participants were chosen through stratified cluster sampling; primary sampling units were chosen from the results of the annual Population and Housing Census, from which representative households were drawn (\cite{O}).
	
	\subsection{Factors/Variables}

	\subsubsection{Cancer Screening Participation}
	
	The main variable of interest is whether an individual has participated in cancer screening in the past 2 years. Possible responses to this question are "yes" and "no."
	
	\subsubsection{Level 1 Factors}
	
	Based on the discussion from section 2, several factors are chosen for this study: education level, income level, exercise frequency, drinking frequency, smoking frequency, whether dietary/nutritional supplements are  taken, nutrition education status and nutrition facts label use. 
	
	Education level is classified into 4 categories: elementary school graduates or below, middle school graduates, high school graduates, and university graduates or higher. Income level is classified into 4 categories as well: low, lower-middle, upper-middle, and high.
	
	Exercise frequency, drinking frequency, and smoking frequency are all a combination of 2 variables from the dataset, and they are classified into 3 categories. Exercise frequency is classified by often (2-7 times a week), occasionally (1-2 times a week), and never/not regularly. Drinking frequency is classified by often (2 or more times a week), occasionally (under 5 times a month), and never/not in the past year. Smoking frequency is classified by daily, occasionally/in the past, and never/less than 5 packs in lifetime.
	
	Whether dietary/nutritional supplements are taken is a response to the question, "have you taken supplements regularly for at least 2 weeks in the past year?," and possible responses are "yes" and "no". Nutrition education status and nutritional facts label use are both binary variables with "yes" and "no" for possible responses.  
	
	\subsubsection{Level 2 Factors: socioeconomic status}
	
	2 univariate and 1 composite socioeconomic status indices are developed. The univariate indices involve one's education level (classified into 4 categories) or income level (classified into 4 categories), averaged across each primary survey unit. The composite index combines the 2 univariate indices by adding them together for each primary survey unit.
	
	To accurately capture differences in each primary survey unit, averaged values were kept as is (instead of creating categories based on a range of numbers).  
	
	\subsubsection{Continuous Variable}
	
	Note that all variables in the multilevel mediation model are configured to be continuous variables. Though certain variables are categorical by nature, to aid the interpretation of results (especially given the large number of variables and categories), all variables are made continuous. 
	
	\subsection{Preliminary Statistical Analysis}

	\subsubsection{Chi-square test}
	
	We first conduct Chi-square tests to assess the chosen factors and determine which are significantly associated with cancer screening participation. 
	
	\begin{table}[h!]
		\centering
		\begin{tabular}{|l|l|l|l|} 
			\hline
			Variable & Degree of Freedom & Chi-squared value & P-value \\ [0.5ex] 
			\hline\hline
			edu & 3 & 147.5246 & 0.000 \\
			\hline
			incm & 3 & 199.9009 & 0.000 \\
			\hline
			nutri\_edu & 1 & 32.7320 & 0.000 \\
			\hline
			diet\_suppl & 1 & 271.6104   & 0.000 \\
			\hline
			label use new & 1 & 2.3040 & 0.129 \\
			\hline
			exer freq & 2 & 11.6863  &  0.003 \\
			\hline
			smk freq & 2 & 323.9205 & 0.000 \\
			\hline
			drink freq & 2 &  34.9039 & 0.000 \\ [1ex] 
			\hline
		\end{tabular}
		\caption{Chi-squared tests with Cancer Screening variable}
		\label{table1}
		\label{Table: 1}
	\end{table}

	Though 1 variable appears to be statistically insignificant with uni-variate analysis, we choose to keep all variables for the multilevel mediation model (all variables are of theoretical importance).
	
	\subsubsection{ICC Analysis}
	
	We calculate intraclass correlation coefficients in mixed-effect models to identify clustering in primary survey units for education levels and income levels.  
	
	\begin{table}[h!]
		\centering
		\begin{tabular}{|l|l|l|l|} 
			\hline
			Variable & Level 2& Level 2 Predictor & ICC\\ [0.5ex] 
			\hline\hline
			Education Level & primary survey unit & NA & .0879755 \\
			\hline
			Income Level & primary survey unit & NA & .2242859 \\ 
			\hline
			Education Level & primary survey unit (psu) & average education level per psu  & 4.52e-23 \\
			\hline
			Income Level & primary survey unit (psu) & average income level per psu & 7.97e-23 \\  [1ex] 
			\hline
		\end{tabular}
		\caption{Intraclass correlation coefficients }
		\label{table1}
		\label{Table: 1}
	\end{table}
	
	Clearly, the addition of level 2 predictors (with random intercepts) significantly reduce the amount of correlation within groups. As such, level 2 predictors will be included in the multilevel mediation model.
	
	\newpage
		
	\subsection{Multilevel Mediation Model}
	
	With the above variables, we now propose the following multilevel mediation model. 
	
	\begin{figure}[h]
		\caption{Multilevel Mediation Model}
		\includegraphics[width = 13 cm]{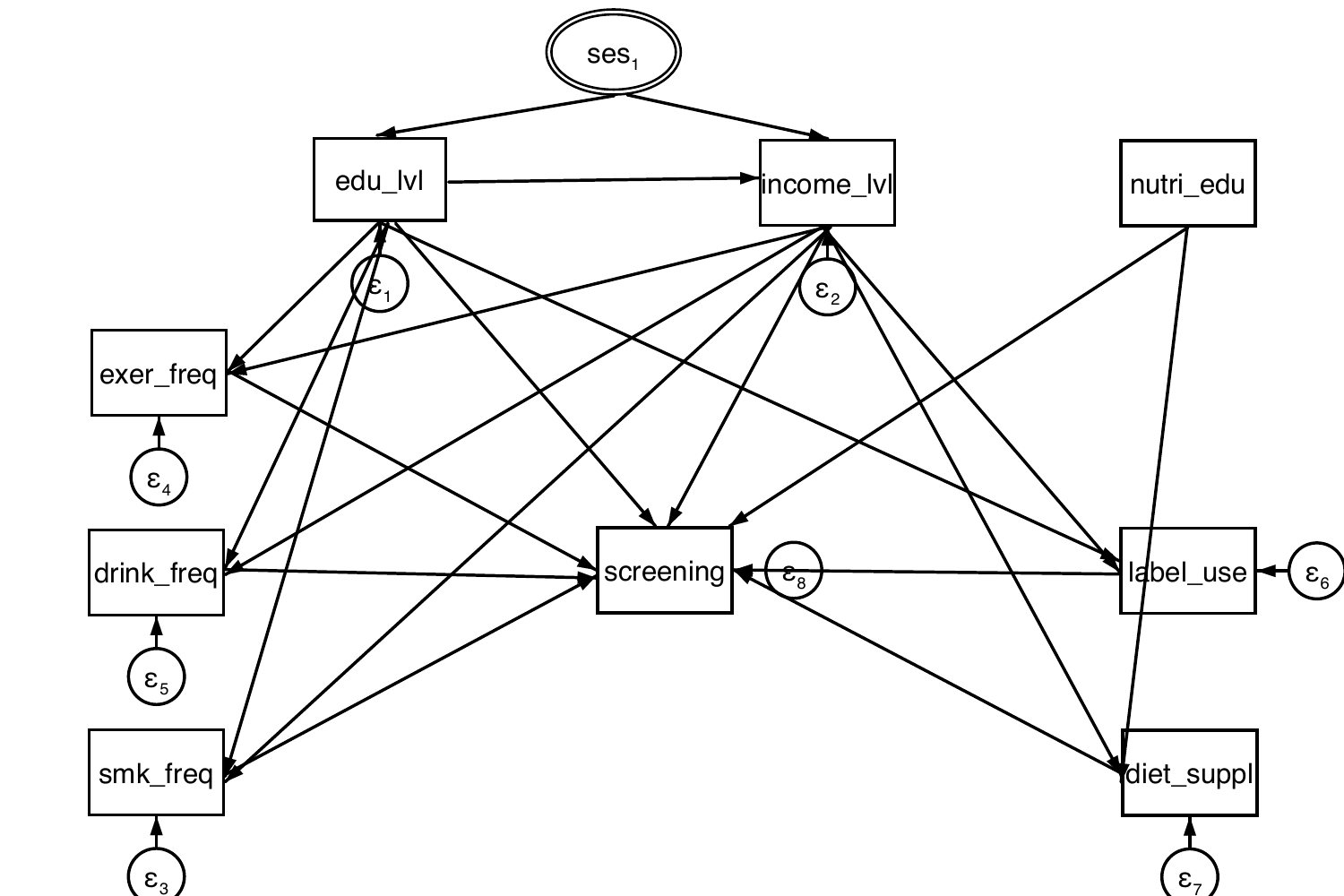}
		\label{Fig1}
	\end{figure}
	$\\$
	Tables 3 and 4 contain the results of the model. Note that the model was run on Stata/IC 16.1
	
	\newpage
	\def\sym#1{\ifmmode^{#1}\else\(^{#1}\)\fi}
\begin{table}[h!]
\centering
\caption{Multilevel Mediation Model: Coefficients}
\resizebox{!}{11cm}{\begin{tabular}{l*{3}{c}}
\hline\hline
            &\multicolumn{1}{c}{(1)}&\multicolumn{1}{c}{(2)}&\multicolumn{1}{c}{(3)}\\
            &\multicolumn{1}{c}{avg\_income\_lvl}&\multicolumn{1}{c}{avg\_edu\_lvl}&\multicolumn{1}{c}{composite\_ses}\\
\hline
screening  &                     &                     &                     \\
exer\_freq   &     0.00831         &     0.00831         &     0.00831         \\
            &      (1.42)         &      (1.42)         &      (1.42)         \\
[1em]
drink\_freq  &     -0.0184\sym{**} &     -0.0184\sym{**} &     -0.0184\sym{**} \\
            &     (-2.61)         &     (-2.61)         &     (-2.61)         \\
[1em]
smk\_freq    &     -0.0768\sym{***}&     -0.0768\sym{***}&     -0.0768\sym{***}\\
            &    (-11.43)         &    (-11.43)         &    (-11.43)         \\
[1em]
label\_use&     -0.0275\sym{**} &     -0.0275\sym{**} &     -0.0275\sym{**} \\
            &     (-2.84)         &     (-2.84)         &     (-2.84)         \\
[1em]
income\_lvl  &      0.0504\sym{***}&      0.0504\sym{***}&      0.0504\sym{***}\\
            &     (12.19)         &     (12.19)         &     (12.19)         \\
[1em]
diet\_suppl&       0.123\sym{***}&       0.123\sym{***}&       0.123\sym{***}\\
            &     (13.45)         &     (13.45)         &     (13.45)         \\
[1em]
edu\_lvl     &     -0.0434\sym{***}&     -0.0434\sym{***}&     -0.0434\sym{***}\\
            &     (-8.74)         &     (-8.74)         &     (-8.74)         \\
[1em]
nutri\_edu&      0.0963\sym{***}&      0.0963\sym{***}&      0.0963\sym{***}\\
            &      (4.98)         &      (4.98)         &      (4.98)         \\
[1em]
\_cons      &       0.664\sym{***}&       0.664\sym{***}&       0.664\sym{***}\\
            &     (28.80)         &     (28.80)         &     (28.80)         \\
\hline
exer\_freq   &                     &                     &                     \\
income\_lvl  &      0.0598\sym{***}&      0.0598\sym{***}&      0.0598\sym{***}\\
            &     (12.13)         &     (12.13)         &     (12.13)         \\
[1em]
edu\_lvl     &       0.121\sym{***}&       0.121\sym{***}&       0.121\sym{***}\\
            &     (24.93)         &     (24.93)         &     (24.93)         \\
[1em]
\_cons      &       0.900\sym{***}&       0.900\sym{***}&       0.900\sym{***}\\
            &     (51.70)         &     (51.70)         &     (51.70)         \\
\hline
drink\_freq  &                     &                     &                     \\
income\_lvl  &     -0.0167\sym{***}&     -0.0167\sym{***}&     -0.0167\sym{***}\\
            &     (-3.62)         &     (-3.61)         &     (-3.61)         \\
[1em]
edu\_lvl     &       0.176\sym{***}&       0.176\sym{***}&       0.176\sym{***}\\
            &     (38.58)         &     (38.62)         &     (38.60)         \\
[1em]
\_cons      &       1.448\sym{***}&       1.448\sym{***}&       1.448\sym{***}\\
            &     (89.99)         &     (90.00)         &     (90.03)         \\
\hline
smk\_freq    &                     &                     &                     \\
income\_lvl  &     -0.0597\sym{***}&     -0.0597\sym{***}&     -0.0597\sym{***}\\
            &    (-11.65)         &    (-11.62)         &    (-11.64)         \\
[1em]
edu\_lvl     &      0.0484\sym{***}&      0.0484\sym{***}&      0.0484\sym{***}\\
            &      (9.34)         &      (9.35)         &      (9.34)         \\
[1em]
\_cons      &       1.346\sym{***}&       1.346\sym{***}&       1.346\sym{***}\\
            &     (73.85)         &     (73.87)         &     (73.88)         \\
\hline
label\_use&                     &                     &                     \\
income\_lvl  &    -0.00308         &    -0.00308         &    -0.00308         \\
            &     (-0.84)         &     (-0.84)         &     (-0.84)         \\
[1em]
edu\_lvl     &      0.0774\sym{***}&      0.0775\sym{***}&      0.0775\sym{***}\\
            &     (21.24)         &     (21.25)         &     (21.25)         \\
[1em]
\_cons      &       0.119\sym{***}&       0.119\sym{***}&       0.119\sym{***}\\
            &      (8.62)         &      (8.62)         &      (8.62)         \\
\hline
income\_lvl  &                     &                     &                     \\
edu\_lvl     &      0.0319\sym{***}&      0.0296\sym{***}&      0.0272\sym{***}\\
            &      (5.70)         &      (5.15)         &      (4.82)         \\
[1em]
M3[avg\_income\_lvl]&           1         &                     &                     \\
            &         (.)         &                     &                     \\
[1em]
M3[avg\_edu\_lvl]&                     &           1         &                     \\
            &                     &         (.)         &                     \\
[1em]
M3[composite\_ses]&                     &                     &           1         \\
            &                     &                     &         (.)         \\
[1em]
\_cons      &       2.379\sym{***}&       2.368\sym{***}&       2.382\sym{***}\\
            &     (80.06)         &     (82.18)         &     (81.73)         \\
\hline
diet\_suppl&                     &                     &                     \\
income\_lvl  &      0.0427\sym{***}&      0.0427\sym{***}&      0.0427\sym{***}\\
            &     (12.81)         &     (12.80)         &     (12.80)         \\
[1em]
nutri\_edu&      0.0320\sym{**} &      0.0320\sym{**} &      0.0320\sym{**} \\
            &      (2.61)         &      (2.61)         &      (2.61)         \\
[1em]
\_cons      &       0.387\sym{***}&       0.387\sym{***}&       0.387\sym{***}\\
            &     (41.94)         &     (41.93)         &     (41.93)         \\
\hline
edu\_lvl     &                     &                     &                     \\
M3[avg\_income\_lvl]&       0.539\sym{***}&                     &                     \\
            &     (30.25)         &                     &                     \\
[1em]
M3[avg\_edu\_lvl]&                     &       0.662\sym{***}&                     \\
            &                     &     (32.09)         &                     \\
[1em]
M3[composite\_ses]&                     &                     &       0.582\sym{***}\\
            &                     &                     &     (31.66)         \\
[1em]
\_cons      &       2.483\sym{***}&       2.466\sym{***}&       2.476\sym{***}\\
            &    (156.12)         &    (137.84)         &    (150.27)         \\
\hline
nutri\_edu&                     &                     &                     \\
M3[avg\_income\_lvl]&      0.0166\sym{***}&                     &                     \\
            &      (3.85)         &                     &                     \\
[1em]
M3[avg\_edu\_lvl]&                     &      0.0150\sym{**} &                     \\
            &                     &      (3.19)         &                     \\
[1em]
M3[composite\_ses]&                     &                     &      0.0165\sym{***}\\
            &                     &                     &      (3.77)         \\
[1em]
\_cons      &       0.101\sym{***}&       0.101\sym{***}&       0.101\sym{***}\\
            &     (46.12)         &     (46.08)         &     (46.09)         \\
\hline

\hline
\(N\)       &       23919         &       23919         &       23919         \\
\hline\hline
\multicolumn{4}{l}{\footnotesize \textit{t} statistics in parentheses}\\
\multicolumn{4}{l}{\footnotesize \sym{*} \(p<0.05\), \sym{**} \(p<0.01\), \sym{***} \(p<0.001\)}\\
\end{tabular}}
\label{tab:label}
\end{table}

	\def\sym#1{\ifmmode^{#1}\else\(^{#1}\)\fi}
\begin{table}[h!]
\centering
\caption{Multilevel Mediation Model: Odds Ratio}
\resizebox{!}{11cm}{\begin{tabular}{l*{3}{c}}
\hline\hline
            &\multicolumn{1}{c}{(1)}&\multicolumn{1}{c}{(2)}&\multicolumn{1}{c}{(3)}\\
            &\multicolumn{1}{c}{avg\_income\_lvl}&\multicolumn{1}{c}{avg\_edu\_lvl}&\multicolumn{1}{c}{composite\_ses}\\
\hline
screening  &                     &                     &                     \\
exer\_freq   &       1.008         &       1.008         &       1.008         \\
            &      (1.42)         &      (1.42)         &      (1.42)         \\
[1em]
drink\_freq  &       0.982\sym{**} &       0.982\sym{**} &       0.982\sym{**} \\
            &     (-2.61)         &     (-2.61)         &     (-2.61)         \\
[1em]
smk\_freq    &       0.926\sym{***}&       0.926\sym{***}&       0.926\sym{***}\\
            &    (-11.43)         &    (-11.43)         &    (-11.43)         \\
[1em]
label\_use&       0.973\sym{**} &       0.973\sym{**} &       0.973\sym{**} \\
            &     (-2.84)         &     (-2.84)         &     (-2.84)         \\
[1em]
income\_lvl  &       1.052\sym{***}&       1.052\sym{***}&       1.052\sym{***}\\
            &     (12.19)         &     (12.19)         &     (12.19)         \\
[1em]
diet\_suppl&       1.131\sym{***}&       1.131\sym{***}&       1.131\sym{***}\\
            &     (13.45)         &     (13.45)         &     (13.45)         \\
[1em]
edu\_lvl     &       0.958\sym{***}&       0.958\sym{***}&       0.958\sym{***}\\
            &     (-8.74)         &     (-8.74)         &     (-8.74)         \\
[1em]
nutri\_edu&       1.101\sym{***}&       1.101\sym{***}&       1.101\sym{***}\\
            &      (4.98)         &      (4.98)         &      (4.98)         \\
\hline
exer\_freq   &                     &                     &                     \\
income\_lvl  &       1.062\sym{***}&       1.062\sym{***}&       1.062\sym{***}\\
            &     (12.13)         &     (12.13)         &     (12.13)         \\
[1em]
edu\_lvl     &       1.128\sym{***}&       1.128\sym{***}&       1.128\sym{***}\\
            &     (24.93)         &     (24.93)         &     (24.93)         \\
\hline
drink\_freq  &                     &                     &                     \\
income\_lvl  &       0.983\sym{***}&       0.983\sym{***}&       0.983\sym{***}\\
            &     (-3.62)         &     (-3.61)         &     (-3.61)         \\
[1em]
edu\_lvl     &       1.192\sym{***}&       1.192\sym{***}&       1.192\sym{***}\\
            &     (38.58)         &     (38.62)         &     (38.60)         \\
\hline
smk\_freq    &                     &                     &                     \\
income\_lvl  &       0.942\sym{***}&       0.942\sym{***}&       0.942\sym{***}\\
            &    (-11.65)         &    (-11.62)         &    (-11.64)         \\
[1em]
edu\_lvl     &       1.050\sym{***}&       1.050\sym{***}&       1.050\sym{***}\\
            &      (9.34)         &      (9.35)         &      (9.34)         \\
\hline
label\_use&                     &                     &                     \\
income\_lvl  &       0.997         &       0.997         &       0.997         \\
            &     (-0.84)         &     (-0.84)         &     (-0.84)         \\
[1em]
edu\_lvl     &       1.081\sym{***}&       1.081\sym{***}&       1.081\sym{***}\\
            &     (21.24)         &     (21.25)         &     (21.25)         \\
\hline
income\_lvl  &                     &                     &                     \\
edu\_lvl     &       1.032\sym{***}&       1.030\sym{***}&       1.028\sym{***}\\
            &      (5.70)         &      (5.15)         &      (4.82)         \\
[1em]
M3[avg\_income\_lvl]&       2.718         &                     &                     \\
            &         (.)         &                     &                     \\
[1em]
M3[avg\_edu\_lvl]&                     &       2.718         &                     \\
            &                     &         (.)         &                     \\
[1em]
M3[composite\_ses]&                     &                     &       2.718         \\
            &                     &                     &         (.)         \\
\hline
diet\_suppl&                     &                     &                     \\
income\_lvl  &       1.044\sym{***}&       1.044\sym{***}&       1.044\sym{***}\\
            &     (12.81)         &     (12.80)         &     (12.80)         \\
[1em]
nutri\_edu&       1.033\sym{**} &       1.033\sym{**} &       1.033\sym{**} \\
            &      (2.61)         &      (2.61)         &      (2.61)         \\
\hline
edu\_lvl     &                     &                     &                     \\
M3[avg\_income\_lvl]&       1.714\sym{***}&                     &                     \\
            &     (30.25)         &                     &                     \\
[1em]
M3[avg\_edu\_lvl]&                     &       1.939\sym{***}&                     \\
            &                     &     (32.09)         &                     \\
[1em]
M3[composite\_ses]&                     &                     &       1.790\sym{***}\\
            &                     &                     &     (31.66)         \\
\hline
nutri\_edu&                     &                     &                     \\
M3[avg\_income\_lvl]&       1.017\sym{***}&                     &                     \\
            &      (3.85)         &                     &                     \\
[1em]
M3[avg\_edu\_lvl]&                     &       1.015\sym{**} &                     \\
            &                     &      (3.19)         &                     \\
[1em]
M3[composite\_ses]&                     &                     &       1.017\sym{***}\\
            &                     &                     &      (3.77)         \\
\hline

\hline
\(N\)       &       23919         &       23919         &       23919         \\
\hline\hline
\multicolumn{4}{l}{\footnotesize Exponentiated coefficients; \textit{t} statistics in parentheses}\\
\multicolumn{4}{l}{\footnotesize \sym{*} \(p<0.05\), \sym{**} \(p<0.01\), \sym{***} \(p<0.001\)}\\
\end{tabular}}
\label{tab:label}
\end{table}

	\section{Results}
	
	We see in all 3 versions of the model that drinking frequency (Odds Ratio = 0.982) and smoking frequency (Odds Ratio = 0.926) are both negatively correlated with cancer screening participation.
	
	We also observe that income level (Odds Ratio = 1.052) is positively correlated with cancer screening participation. Nutrition education status (Odds Ratio = 1.101) and the consumption of dietary supplements (Odds Ratio = 1.131) are shown to be positively associated with cancer screening participation as well.
	
	Furthermore, education level, nutrition education status, and income level appear to have significant effects on other variables. Income level is positively correlated with exercise frequency (Odds Ratio = 1.062) and negatively correlated with drinking frequency (Odds Ratio = 0.983) and smoking frequency (Odds Ratio = 0.942). On the other hand, education level is positively correlated with exercise frequency (Odds Ratio = 1.128), drinking frequency (Odds Ratio = 1.192), smoking frequency (Odds Ratio = 1.050), and income level (Odds Ratio = 1.032). 
	
	Moreover, an individual's income level (Odds Ratio = 1.044) and nutrition education status (Odds Ratio = 1.033) are both positively correlated with the use of diet supplements. Education level is shown to be positively associated with nutrition label usage as well (Odds Ratio = 1.081).   
	
	Finally, we see that all variations of the level 2 predictor (socioeconomic status index) are significantly associated with an individual's education level (Odds Ratio with average income level: 1.714; average education level: 1.939; composite index: 1.790) and nutrition education status (Odds Ratio with average income level: 1.017; averaged education level: 1.015; composite index: 1.017).

	We now address some unexpected results of the proposed model. Education level was found to be negatively correlated with cancer screening participation (Odds Ratio: 0.958). To examine the effect of education level more closely, we conduct logistic regression with a categorical education level variable of greater specificity (7 categories).
	
	\def\sym#1{\ifmmode^{#1}\else\(^{#1}\)\fi}
\begin{table}[h!]
\parbox{.45\linewidth}{
\centering

\begin{tabular}{l*{1}{c}}
	\hline\hline
	&\multicolumn{1}{c}{(1)}\\
	&\multicolumn{1}{c}{screening}\\
	\hline
	screening  &                     \\
	No Education  &           0         \\
	&         (.)         \\
	[1em]
	Elementary School  &       0.698\sym{***}\\
	&      (7.83)         \\
	[1em]
	Middle School  &       0.829\sym{***}\\
	&      (8.88)         \\
	[1em]
	High School  &       0.539\sym{***}\\
	&      (6.40)         \\
	[1em]
	2/3 Year College  &     -0.0449         \\
	&     (-0.51)         \\
	[1em]
	4 Year College  &     -0.0189         \\
	&     (-0.22)         \\
	[1em]
	Graduate School &       0.636\sym{***}\\
	&      (6.12)         \\
	[1em]
	\_cons      &      0.0277         \\
	&      (0.35)         \\
	\hline
	\(N\)       &       17564         \\
	\hline\hline
	\multicolumn{2}{l}{\footnotesize \textit{t} statistics in parentheses}\\
	\multicolumn{2}{l}{\footnotesize \sym{*} \(p<0.05\), \sym{**} \(p<0.01\), \sym{***} \(p<0.001\)}\\
\end{tabular}
\caption{Logistic Regression: Coefficients}
}
\hfill
\parbox{.45\linewidth}{
\centering
\begin{tabular}{l*{1}{c}}
	\hline\hline
	&\multicolumn{1}{c}{(1)}\\
	&\multicolumn{1}{c}{screening}\\
	\hline
	screening &                     \\
	No Education  &           1         \\
	&         (.)         \\
	[1em]
	Elementary School  &       2.010\sym{***}\\
	&      (7.83)         \\
	[1em]
	Middle School  &       2.290\sym{***}\\
	&      (8.88)         \\
	[1em]
	High School  &       1.714\sym{***}\\
	&      (6.40)         \\
	[1em]
	2/3 Year College  &       0.956         \\
	&     (-0.51)         \\
	[1em]
	4 Year College  &       0.981         \\
	&     (-0.22)         \\
	[1em]
	Graduate School  &       1.889\sym{***}\\
	&      (6.12)         \\
	\hline
	\(N\)       &       17564         \\
	\hline\hline
	\multicolumn{2}{l}{\footnotesize Exponentiated coefficients; \textit{t} statistics in parentheses}\\
	\multicolumn{2}{l}{\footnotesize \sym{*} \(p<0.05\), \sym{**} \(p<0.01\), \sym{***} \(p<0.001\)}\\
\end{tabular}
\caption{Logistic Regression: Odds Ratio}
}
\end{table}
	
	We see that compared to lack of schooling, having attended elementary school, middle school, high school, or graduate school all play a significant role (positive correlation) in cancer screening participation. However, having attended high school or college is not correlated with cancer screening participation ($p \geq 0.05$)--indicative of why the direct effect of education on cancer screening participation is "diluted" in the multilevel mediation model.
	
	In addition, nutrition label usage is found to be negatively correlated with cancer screening participation (Odds Ratio: 0.973). This may be caused by the fact that nutrition label usage has a statistically insignificant relationship with cancer screening participation when univariate analysis is done (Chi-square test); hence, a suppression effect.
	
	As a whole, we see that education level, income level, and nutrition education status play a key role in cancer screening participation. These 3 variables both directly affect cancer screening participation and are significantly associated with other relevant factors.
	
	In particular, nutrition education status has the second largest direct effect on cancer screening participation and is positively correlated with the largest direct effect (use of diet supplements).
	
	Taking a closer look at the 3 socioeconomic status indices, we see that the composite index is most applicable to this model given its high statistical significance ($p < 0.001$) for education level, nutrition education status and income level and its larger odds ratio value relative to the other indices.
	
	Note that the coefficient of the path from socioeconomic status indices to income level is constrained to 1 (with odds ratio 2.718) for all 3 variations.

	\section{Discussion}
	
	Consistent with results from past studies, drinking frequency and smoking frequency are shown to be negatively correlated with cancer screening participation (\cite{A}). These 2 factors are generally considered causes or risk factors for certain cancers. Hence, smokers and drinkers would benefit from increased cancer screening test participation, especially for lung cancer, larynx cancer, throat cancer, and other high-risk cancers. Given the large number of smokers and drinkers in Korea, cancer screening campaigns focused on such groups should be set up by health authorities. 
	
	Also, like past studies, income in found to be positively correlated with cancer screening participation (\cite{A}). Even with national health insurance and subsidized screening costs, income is still a relevant factor. Possible causes include the following: not all screening tests are offered at lowered costs and higher income often results in more available time for health check-ups.     
	
	As past studies have shown, dietary factors (use of dietary supplements) are positively correlated with cancer screening participation  (\cite{A}). This can be explained by the fact that individuals who care about their health are willing to participate in regular check-ups.   
	
	However, statistically significant relationships between exercise frequency and cancer screening participation are not found in this study; contrary to past studies (\cite{B}). Unlike dietary factors, exercise is easily accessible to most individuals; therefore, correlations are less likely to be observed. 
	
	Furthermore, the relationship between education levels and exercise frequency, drinking frequency, and dietary factors (use of nutrition labels) are in line with previous studies (\cite{C}, \cite{D}). Also, the relationship between education levels and income levels is consistent with past surveys (\cite{J}). However, unlike past research, smoking frequency is found to have positive associations with education (\cite{C}). This can be explained simply by the prevalence of smoking in Korea (\cite{P}). To reduce smoking rates, health authorities should increase smoking campaigns, especially in high schools around the nation. 
	
	Looking at income, the relationship between income levels and exercise frequency, dietary factors (use of dietary supplements) and smoking frequency is consistent with past studies (\cite{E}, \cite{F},\cite{G}). However, unlike past studies, drinking frequency is shown to have negative associations with income (\cite{E}). Increased awareness of the risks of heavy drinking may be related with these results. To reduce heavy drinking, health authorities should consider increasing health campaigns focused on lower income regions.
	
	The relationship between nutrition education status and dietary supplement usage is consistent with previous findings as well (\cite{I}). 
	
	However, no statistically significant relationship between income and nutrition label use was found. We believe that this may be related with the fact that individuals of lower income need to plan out their meals, often with nutrition in mind.
	
	Taking a look at education levels, we see that its effect on cancer screening participation are contrary to previous studies--even with univariate logistic regression, we find that having attended college or high school is not correlated with cancer screening participation (\cite{A}). This can be explained by the prevalence of nutrition education programs that do not require higher levels of education. With such programs, individuals can realize the need for and availability of cancer screening tests by visiting local public health centers, community centers and welfare facilities. Given the significance of nutrition education status (as shown with its high odds ratio) with cancer screening participation, local health authorities should consider expanding the availability of these programs.
	
	Finally, because regional socioeconomic status factors are found to be significantly correlated with education levels, nutrition education statuses, and income levels (variables which are associated with other relevant variables as well), improving access to education opportunities and nutritional education centers should be prioritized by local governments. $\\$
	
	Overall, we found in this study that education and income play a key role in cancer screening participation. Our findings are significant in that we expanded upon previous research, identifying the role of education and income not only on cancer screening participation but also on other relevant factors through a multilevel mediation model (with considerations to regional socioeconomic status indices).
	
	We were also successful in developing a composite socioeconomic status index which accurately captured  differences in each primary survey unit for education and income.  
	
	Nevertheless, the study has some limitations. First, the data obtained from the Korea National Health and Nutrition Examination Survey was, for the most part, a self-report study. Thus, the data used may have contained some inaccuracies which could've impacted the results of the model. Second, as is the nature of cross-sectional data, we cannot establish causal relationships or infer causality with our mediation model.
	
	We hope that related studies continue in the future, expanding upon our results and improving on aforementioned flaws. Causal research, for instance, could help determine causality. Another option would be to incorporate more variables like working hours and distance to public health centers into the mediation model. 
	
	Even with these limitations, our novel findings and policy recommendations (special attention should be paid to increasing accessibility of nutrition education programs) may be of use for health authorities aiming to increase participation in cancer screening tests.

	\bibliographystyle{amsplain}

\end{document}